\documentclass{ws-procs9x6}

\begin{document}

\title{SIGNATURES OF DARK MATTER ANNIHILATION IN THE LIGHT OF PAMELA/ATIC ANOMALY}

\author{KAZUNORI NAKAYAMA}

\address{Institute for Cosmic Ray Research, University of Tokyo, \\
Kashiwa, Chiba 277-8582, Japan\\
E-mail: nakayama@icrr.u-tokyo.ac.jp}

\begin{abstract}
Recent measurements of cosmic-ray electron and positron fluxes by
  PAMELA and ATIC experiments may indicate the existence of
  annihilating dark matter with large annihilation cross section.
  We discuss its possible relation to other astrophysical/cosmological observations :  
  gamma-rays, neutrinos, and big-bang nucleosynthesis.
  It is shown that they give stringent constraints on some annihilating dark matter models.
\end{abstract}


\bodymatter

\section{Introduction} \label{sec:intro}

Recently, the PAMELA satellite experiment \cite{Adriani:2008zr}
reported an excess of the cosmic ray positron flux
above the energy $\sim 10$~GeV, and the ATIC balloon experiment \cite{:2008zz} also
found an excess of the electron plus positron flux, whose peak energy is around 600~GeV.
These results are now drawing lots of attention of particle physicists,
since they can be interpreted as signatures of dark matter (DM) annihilation/decay.
Although some astrophysical sources are proposed \cite{Hooper:2008kg},
we focus on the DM annihilation scenario as an explanation of the PAMELA/ATIC anomaly
\cite{Bergstrom:2008gr}.

It is known that if the positron/electron excess is caused by the DM annihilation,
a large annihilation cross section $\langle \sigma v \rangle \sim 10^{-24}$-$10^{-23}$
$~{\rm cm^3s^{-1}}$,
which is two or three magnitude of larger than the standard value
for reproducing the observed DM abundance in the thermal relic scenario, is needed.
Sommerfeld enhancement mechanism may provide a large annihilation rate \cite{Hisano:2003ec}.
Although a DM with large annihilation cross section does not lead to correct DM abundance,
DM can also be produced nonthermally, through the decay of long-lived particles, for example
\cite{Kawasaki:1995cy}.
Thus non-thermal DM is a good candidate as a source of the PAMELA/ATIC anomaly.
In this study, we show that those DM models are constrained from other observations :
gamma-rays, high-energy neutrinos and big-bang nucleosynthesis (BBN).\footnote{
	One can avoid these constraints,
	if possible clumpy structures of the DM halo in the Galaxy are taken into account.
	However, such small scale structures of the DM halo are not well understood yet.
	In the following, we assume smooth distribution of DM halo.
}
This study is based on the works with J.~Hisano, M.~Kawasaki, K.~Kohri and T.~Moroi
\cite{Hisano:2008ti,Hisano:2009rc,Hisano:2008ah}.

\section{Cosmic Rays from Dark Matter Annihilation}

\subsection{Positrons and Electrons}

\begin{figure}
\begin{center}
\psfig{file=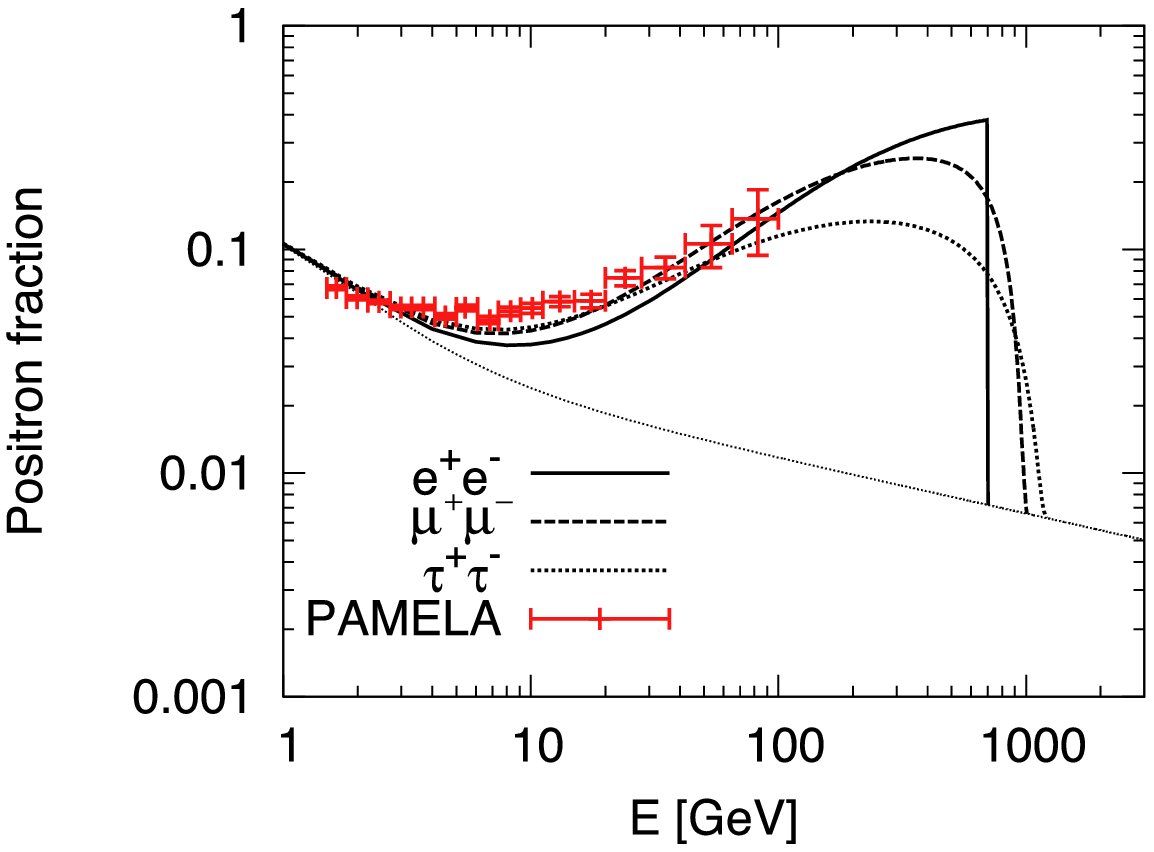,width=3.0in}
\psfig{file=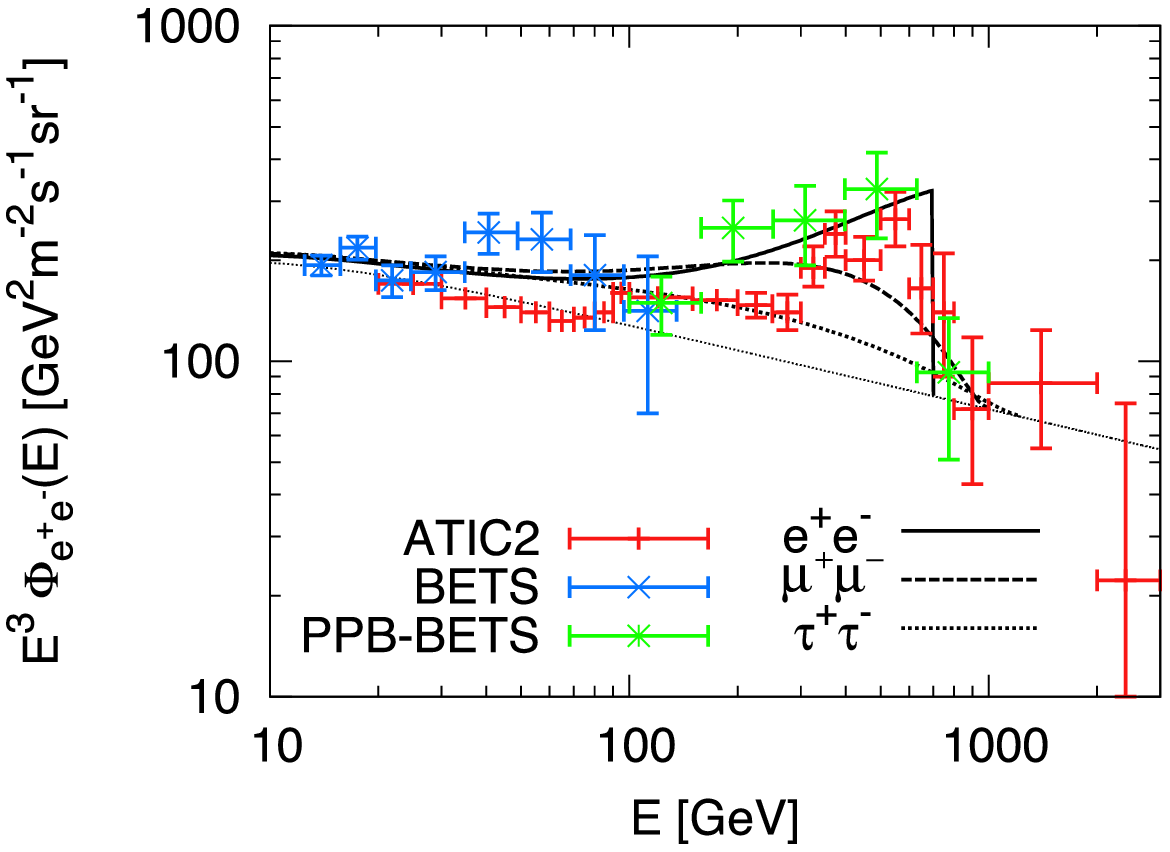,width=3.0in}
\end{center}
\caption{(Top) The ratio of $e^+$ to $e^+ + e^-$ flux as a function of its energy, for the case of
DM annihilating into $e^+ e^-$, $\mu^+ \mu^-$ and $\tau^+ \tau^-$.
(Bottom) Total $e^+ + e^-$ flux times $E^3$ for the same DM models as top panel.}
\label{fig:pos}
\end{figure}

In the Galaxy, DM annihilates each other producing high energy particles,
including positrons and electrons.
High-energy charged particles propagates under an influence by the
highly tangled magnetic fields.
They also lose their energy during the propagation 
inside the Galaxy due to the synchrotron emission and inverse Compton scattering processes
with CMB photon and diffuse star light.
Thus their motion can be regarded as a random walk process with energy loss,
and described by the following diffusion loss equation \cite{Baltz:1998xv},
\begin{equation}
	\frac{\partial f}{\partial t}(E, \vec x)=
	K(E)\nabla^2 f(E,\vec x) +\frac{\partial}{\partial E}[b(E) f(E,\vec x)] +Q(E,\vec x),
\end{equation}
where $f(E,\vec x)$ is the positron/electron number density with energy $E$ 
at the position $\vec x$, $K(E)$ is the diffusion constant and $b(E)$
denotes the energy loss due to the synchrotron emission and inverse Compton scattering process
off the microwave background photons and diffuse star light.
The source term, $Q(E,\vec x)$, represents the injection from DM annihilation.
Then we can solve the diffusion equation in a semi-analytical way \cite{Hisano:2005ec}.
Results are shown in Fig.~\ref{fig:pos}.
The top panel shows the ratio of the positron to the sum of electron plus positron flux,
and the bottom panel shows the electron plus positron flux multiplied by $E^3$.
We consider three DM models : DM annihilates into $e^+ e^-$ with $m=700$~GeV
and $\langle \sigma v \rangle=5\times 10^{-24}~\rm{cm^3s^{-1}}$, 
into $\mu^+ \mu^-$ with $m=1$~TeV
and $\langle \sigma v \rangle=1.5\times 10^{-23}~\rm{cm^3s^{-1}}$,
and into $\tau^+ \tau^-$ with $m=1.2$~TeV
and $\langle \sigma v \rangle=2\times 10^{-23}~\rm{cm^3s^{-1}}$.
Background flux is taken from the simulation by Moskalenko and Strong \cite{Moskalenko:1997gh}.
Data points of PAMELA \cite{Adriani:2008zr}, ATIC \cite{:2008zz}, 
BETS \cite{Torii:2001aw} and PPB-BETS \cite{Torii:2008xu} are also plotted.
It is seen that these DM models well reproduce the PAMELA/ATIC anomaly.
Notice that no anti-protons are produced in these models.

\subsection{Gamma-Rays}

\begin{figure}[t]
\begin{center}
\psfig{file=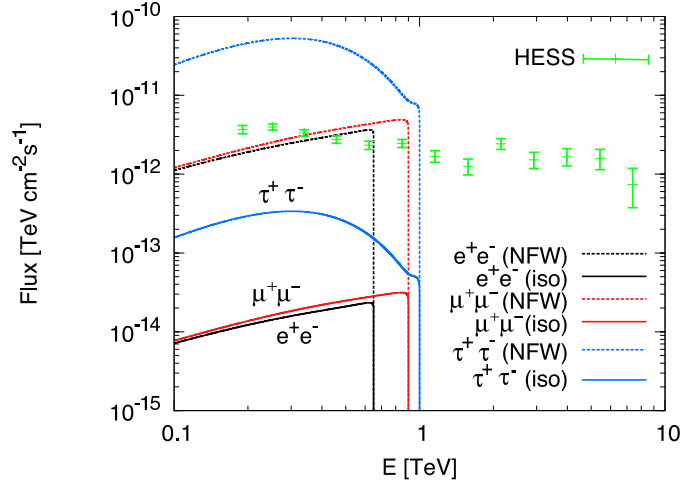,width=3.5in}
\end{center}
\caption{Gamma-ray flux from the Galactic center for the DM models used in Fig.~\ref{fig:pos},
both for NFW and isothermal density profiles.}
\label{fig:gam}
\end{figure}

We have seen that the DM annihilation model reproducing the PAMELA/ATIC anomaly requires
large annihilation cross section,  $\langle \sigma v \rangle \sim10^{-23}~\rm{cm^3s^{-1}}$.
Such models would also produce significant gamma-ray flux from the Galactic center.
which should be compared with HESS observations \cite{Aharonian:2004wa}.
As opposed to the case of electrons/positrons, gamma-rays do not suffer from energy loss
and comes to the Earth in a straight way.
Thus the evaluation of the gamma-ray flux is rather simple, but it has large uncertainty 
related to the lack of the understanding of the DM density profile near the Galactic center,
which leads to orders of magnitude uncertainty in the resultant gamma-ray flux.
As examples, we adopt two DM halo models : Navarro-Frenk-White (NFW) model \cite{Navarro:1995iw}
where the DM number density scales as $n_{\rm DM}(r)\sim 1/r$ near the Galactic center
with distance from the Galactic center $r$,
and isothermal profile which scales as $1/(a^2 + r^2)$ with core radius $a$.
Fig.~\ref{fig:gam} shows the gamma-ray flux from the region $0.1^{\circ}$ around the Galactic center
for the DM models used in Fig.~\ref{fig:pos}
with both NFW and isothermal DM density profile.
We have included both contributions from internal bremsstrahlung and cascade decay processes.
It is seen that the cuspy density profile as NFW profile is not favored, otherwise
the gamma-ray flux exceeds the HESS observations.
However, if a more moderate density profile is adopted,
the gamma-ray constraint is satisfied.

\subsection{Neutrinos}

\begin{figure}[t]
\begin{center}
\psfig{file=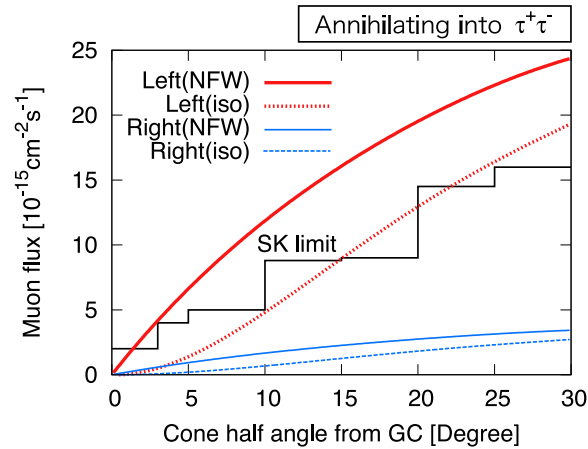,width=3.0in}
\end{center}
\caption{ The up-going muon flux for the DM annihilating into 
left-handed $\tau$ pair ($\tau_L^- \tau_R^+ + \nu_\tau \bar \nu_\tau$)
and right-handed $\tau$ pair ($\tau_R^- \tau_L^+$) for NFW and isothermal profile. 
Limits from SK are also shown.}
\label{fig:taunu}
\end{figure}

Similar to the gamma-rays, high-energy neutrinos are also produced by the DM annihilation,
which can be observed or constrained by on-going or future neutrino detectors
\cite{Beacom:2006tt},
through the Cherenkov light emission by high-energy muons arising from the interaction of
neutrinos with matter inside the Earth.
Interestingly, the Super-Kamiokande (SK) results \cite{Desai:2004pq} already give constraints on
DM models explaining the PAMELA/ATIC anomaly \cite{Hisano:2008ah,Liu:2008ci}.
Fig.~\ref{fig:taunu} shows the resulting up-going muon flux for the case of DM annihilating into
$\tau^+ \tau^-$ with mass and cross sections are fixed to reproduces the PAMELA/ATIC results
\cite{Hisano:2008ah}.
Here we have distinguished two cases : annihilation into 
left-handed $\tau$'s and right-handed $\tau$'s,
since it is natural to expect that the DM also directly annihilates into neutrino pair 
if the final state particles are left-handed.
Thus we consider two cases where the DM annihilates into left-handed leptons
 ($\tau_L^- \tau_R^+ + \nu_\tau \bar \nu_\tau$) and right-handed leptons
 ($\tau_R^- \tau_L^+$).
 It is seen that the former case may conflict with present bound from SK.

\section{Constraints from Big-Bang Nucleosynthesis}

Finally, we discuss constraint on the DM annihilation cross section from the BBN.
If there are some additional energy injection processes during BBN,
light elements such as He, D, Li may be destroyed or overproduced,
which may be in conflict with observations of primordial light element abundances
\cite{Kawasaki:2004yh}.
Some fraction of the DM still annihilates each other even after the freeze-out epoch
where the DM number-to-entropy ratio is fixed,
and such effects can have significant effects on BBN if the annihilation cross section is large enough
\cite{Reno:1987qw}.
Effects of DM annihilation on BBN was discussed by Jedamzik \cite{Jedamzik:2004ip},
with an emphasis on a possible explanation to the cosmic lithium problem.
Here we provide conservative constraints on the DM annihilation rate,
in the case of pure radiative annihilation mode \cite{Hisano:2009rc}.

\begin{figure}[t]
\begin{center}
\psfig{file=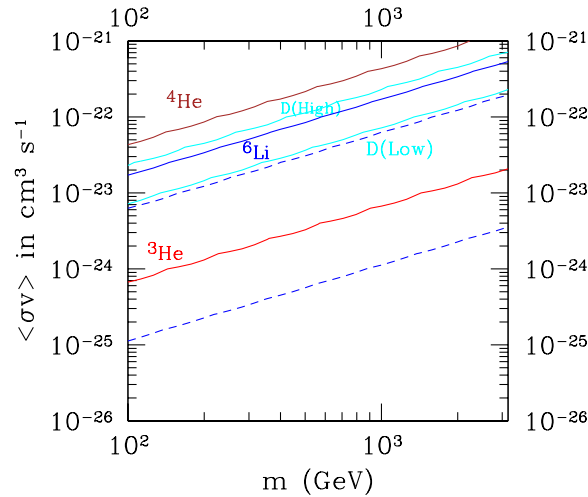,width=3.0in}
\end{center}
\caption{Constraints on the DM annihilation cross section from observations of various 
	light element abundances.
	Here pure radiative annihilation is assumed.
	For reference, the region sandwiched by two dashed lines are allowed from 
	${\rm {^6}Li}$ abundance if no stellar depletion is assumed. }
\label{fig:BBN}
\end{figure}

Fig.~\ref{fig:BBN} shows constraints on the annihilation cross section into radiative mode
from observation of light element abundances.
We adopt the following light element abundances as observed primordial values :
$n_{\rm D}/n_{\rm H}=(2.82\pm 0.26) \times 10^{-5}$ for ``Low'' value of D 
and $n_{\rm D}/n_{\rm H}=(3.98_{-0.67}^{+0.59}) \times 10^{-5}$ for ``High'' value of D \cite{O'Meara:2006mj},
$n_{\rm {^3}He}/n_{\rm D}<1.10$ for ${\rm {^3}He}$ \cite{GG03},
$Y_{\rm p}=0.2516 \pm 0.0040$ for ${\rm {^4}He}$ \cite{Izotov:2007ed,Fukugita:2006xy},
$\log_{10}(n_{\rm {^7}Li}/n_{\rm H})=-9.90\pm 0.09+0.35$ for ${\rm {^7}Li}$
\cite{Bonifacio:2006au,RBOFN},
$n_{\rm {^6}Li}/n_{\rm {^7}Li} < 0.046\pm 0.022 +0.106$ for ${\rm {^6}Li}$ \cite{Asplund:2005yt}
taking into account the effect of rotational mixing in the star \cite{Pinsonneault:2001ub}.
Note that the energy fraction going into the radiation $(f_{\rm vis})$ is 35\% for the case of $\mu$,
and 32\% for the case of $\tau$.
Thus the vertical axis in Fig.~\ref{fig:BBN} should be regarded as a constraint on
$f_{\rm vis} \langle \sigma v \rangle$.
We can see that DM models adopted in Fig.~\ref{fig:pos} are marginally consistent with BBN,
or may be disfavored for the case of annihilating into $\tau$'s.

\section{Conclusions}

Motivated by the recent observations of anomalous cosmic-ray electron/positron fluxes
by PAMELA and ATIC experiments,
we have investigated observational implications of DM annihilation scenario
which account for the electron/positron anomalies
in a model independent way.
Since a large annihilation cross section is required in order to reproduce PAMELA/ATIC anomaly,
it may leave characteristic signatures on other observations :
gamma-rays, neutrinos and primordial light element abundances.
What we have found is that all these observations give stringent constraints on 
DM annihilation models accounting for the PAMELA/ATIC anomaly.
If the density profile of the DM halo in our Galaxy is cuspy, like NFW profile for example,
the HESS results exclude DM annihilation models.
This constraint can be avoided if more moderate density profile is adopted.
Neutrino constraint may also be so severe for the DM annihilating into left-handed leptons
that the up-going muon flux may exceed the SK bound.
The BBN constraint is also so stringent that DM annihilation models are only marginally allowed, 
or some models are already disfavored.
All these possible signatures are important for finding or constraining DM annihilation models
in connection with cosmic positron/electron anomalies.\\

{\it Acknowledgments--} 
I would like to thank J.~Hisano, M.~Kawasaki, K.~Kohri and T.~Moroi
for their collaboration.



\end{document}